\documentclass[12pt,aps,prd,reprint,sort&compress, showpacs, nofootinbib, 
               preprintnumbers]{revtex4-1}
\pdfoutput=1
\usepackage{graphicx}
\usepackage{amsmath}
\usepackage{amssymb}
\usepackage{dsfont}
\usepackage{amsfonts}
\usepackage[UKenglish]{babel}
\usepackage{hyperref}
\usepackage{nicefrac}
\usepackage{verbatim}
\usepackage{bbm}
\usepackage{color}
\usepackage{multirow}
\usepackage{cleveref}
\usepackage{braket}

\begin{document}
\title{Angular momentum content of the $\rho(1450)$ from chiral
lattice fermions} 
 
\author{C.~Rohrhofer}
\email{christian.rohrhofer@uni-graz.at}
\author{M.~Pak}
\email{markus.pak@uni-graz.at}
\author{L.~Ya.~Glozman}
\email{leonid.glozman@uni-graz.at}
\affiliation{Institut f\"ur Physik, FB Theoretische Physik, Universit\"at Graz,
             Universit\"atsplatz 5, 8010 Graz, Austria}
\begin{abstract}
We identify the chiral and angular momentum content of the 
leading quark-antiquark Fock component for the $\rho(770)$ and
$\rho(1450)$ mesons using a two-flavor lattice simulation with dynamical
Overlap Dirac fermions. We extract this information from the overlap factors 
of two  interpolating fields with different chiral structure  and from the unitary transformation
between chiral and angular momentum basis. For the chiral content of the mesons
we find that
the $\rho(770)$ slightly favors the $(1,0)\oplus(0,1)$ chiral representation
and the $\rho(1450)$ slightly favors the $(1/2,1/2)_b$ chiral
representation. In the angular momentum basis the $\rho(770)$ is then a $^3S_1$
state, in accordance with the quark model. The $\rho(1450)$ is a $^3D_1$ state, 
showing that the quark model wrongly assumes the $\rho(1450)$ to be a 
radial excitation of the $\rho(770)$.  
\end{abstract}

\maketitle
\section{Introduction}
The potential constituent quark model has been quite successful in describing the low-lying
hadron spectrum \cite{Agashe:2014kda}. Being an effective classification scheme, it does not care
about foundations in terms of  underlying QCD dynamics. Despite its successes
the non-relativistic description clearly has limitations.

In this paper we investigate the angular momentum content of the $\rho(770)$
and $\rho(1450)$ mesons. In the spectroscopic notation $n\;^{2S+1}l_J$ the
$\rho(770)$ is assigned to the $1\;^3S_1$ state by the quark model. The 
$\rho(1450)$
is assigned to the $2\;^3S_1$ state, hence being the first radial excitation of
the $\rho(770)$. However, this assumption is by far not clear from the
underlying QCD dynamics, and is an output of the non-relativistic potential
description of a meson as a two-body system.

In principle the angular momentum content of the leading quark-antiquark
Fock  components of  mesons can be identified
by a lattice simulation ~\cite{Glozman:2009rn,Glozman:2009cp,Glozman:2010zn,
Glozman:2011gf,Glozman:2011nk}. The crucial ingredients to such a study are the
overlap factors obtained with operators that form a complete set
with respect to the chiral-parity group.
From these overlap factors the chiral content of a state can be
identified. Then, given a unitary transformation between the
chiral basis and the $^{2S+1}l_J$ basis we can reconstruct the angular momentum content.  Since the chiral content is important for such a study we need a
lattice fermion discretization, which respects chiral symmetry. This is
why we use overlap fermions, which distinguishes the present study
from the previous ones.  

We find in contrast to the previous studies that the $\rho(1450)$ is practically
a pure $1\;^3D_1$
state. We will argue that our result is correct due to a careful analysis of
the signs of the overlap factors.

Further, we remove the low-lying Dirac eigenmodes of the spectrum, which has
been done recently  to show an emergent $SU(2N_f)$
symmerty in the QCD spectrum. This symmetry connects all flavors and quark
chiralities, which means in the two-flavor case that $u_L$, $u_R$, $d_L$, $d_R$
are connected with each other. Here we study the effect of the Dirac eigenmode
removal on the overlap factors of \textit{vector} and \textit{pseudotensor}
interpolators.

The outline of the article is as follows: In section \ref{cFormalism} we
present the method how to extract the chiral and angular momentum content of
the physical states. In section \ref{cResolution-scale} we discuss our
simulation parameters and introduce a resolution scale for our measurements.
In section \ref{cResults} we present our main findings. In section
\ref{cUnbreaking} we discuss the effect of removing the low-lying eigenmodes
on the $\rho$ states. Finally, in  section \ref{cConclusions} we give a short
conclusion.

\section{Formalism}
\label{cFormalism}
The formalism how to extract chiral and angular momentum content from lattice
correlators has been explained in detail in Ref.~\cite{Glozman:2009cp}. We
review the basic steps.

To generate states with  $\rho$ quantum numbers $(1,1^{--})$ two different local
interpolators can be used, which belong to two distinct chiral representations
\footnote{For a detailed description of the chiral-parity group we refer
to Ref.~\cite{Glozman:2007ek}.}
\begin{align}
\label{veccurrent}
J_\rho^V(x) = \bar{\Psi}(x)(\tau^a \otimes \gamma^i) \Psi(x)\quad \in (0,1)\oplus(1,0) \\
\label{pseudocurrent}
J_\rho^T(x) = \bar{\Psi}(x)(\tau^a \otimes \gamma^0\gamma^i)\Psi(x) \quad \in (1/2,1/2)_b.
\end{align}
We denote them according to their Dirac structure as \textit{vector (V)} and
\textit{pseudotensor (T)} interpolators. Both of them couple to the physical
$\rho$ states. 
The interpolators (\ref{veccurrent}), (\ref{pseudocurrent})
transform differently under $SU(2)_L \times SU(2)_R$ and therefore belong to two
different chiral representations.
If chiral symmetry would be manifest in nature, these two
interpolators would generate two different particles and the index of the
irreducible representation of the chiral-parity group would be an additional
quantum number. In the real world, where chiral symmetry is broken, a physical
$\rho$-meson is a mixture of two possible chiral representations and consequently both interpolators create the same physical $\rho$-meson.

In a next step we connect the chiral basis to the 
 angular momentum basis with quantum numbers isospin $I$ and
$^{2S+1}l_J$. For spin-$1$ isovector mesons there are only two allowed
states $\ket{1;^3S_1}$ and $\ket{1;^3D_1}$, which are connected to the
chiral basis by a unitary transformation \cite{Glozman:2007at}:
\begin{align}
\label{rhovec}
\ket{\rho_{(0,1)\oplus(1,0)}} &= \sqrt{\frac{2}{3}} \ket{1;^3S_1} +
								    \sqrt{\frac{1}{3}} \ket{1;^3D_1} \;, \\
\label{rhotensor}
\ket{\rho_{(1/2,1/2)_b}} &= \sqrt{\frac{1}{3}} \ket{1;^3S_1} -
							   \sqrt{\frac{2}{3}} \ket{1;^3D_1}\;.
\end{align}
This transformation is valid only in the rest frame.

On the lattice we evaluate the correlators $\braket{J(t)J^\dagger(0)}$.
We apply the variational technique, where different interpolators are used to
construct the correlation matrix $\braket{J_l(t)J^\dagger_m(0)}=C(t)_{lm}$.
By solving the generalized eigenvalue problem
\begin{align}
C(t)_{lm}u_m^{(n)}=\lambda^{(n)}(t,t_0)C(t_0)_{lm}u_m^{(n)}
\end{align}
the masses of the states can be extracted from the eigenvalues:
\begin{align}
\lambda^{(n)}(t,t_0)=e^{-E^{(n)}(t-t_0)}
\left(1+\mathcal{O}\left(e^{-\Delta E^{(n)}(t-t_0)}\right)\right) \;.
\end{align}
A single entry of the correlation matrix reads
\begin{align}
\braket{J_l(t)J_m^\dagger(0)}=\sum_n a_l^{(n)}a_m^{(n)*}e^{-E^{(n)}t} \;,
\end{align}
where $a_l^{(n)}=\bra{0}J_l\ket{n}$ is the overlap of interpolator $J_l$ with
the physical state $\ket{n}$. The ratio of these overlap factors gives a
relative weight  of
the different chiral representations in a given physical state.
It can be constructed as
\begin{align}
\frac{a_l^{(n)}}{a_k^{(n)}}=\frac{C(t)_{lj}u_j^{(n)}}{C(t)_{kj}u_j^{(n)}} \;.
\label{ratio}
\end{align}
We note that the ratio of eigenvector components $u_l$ is not suited due to
the lack of a unique normalization for different operators.

We can extract the ratio $a_V/a_T$ for each state $n$. Then via the unitary
transformation (\ref{rhovec}),(\ref{rhotensor}) we  arrive at
the angular momentum content of the $\rho$ mesons.

\begin{figure}
\includegraphics[scale=0.52]{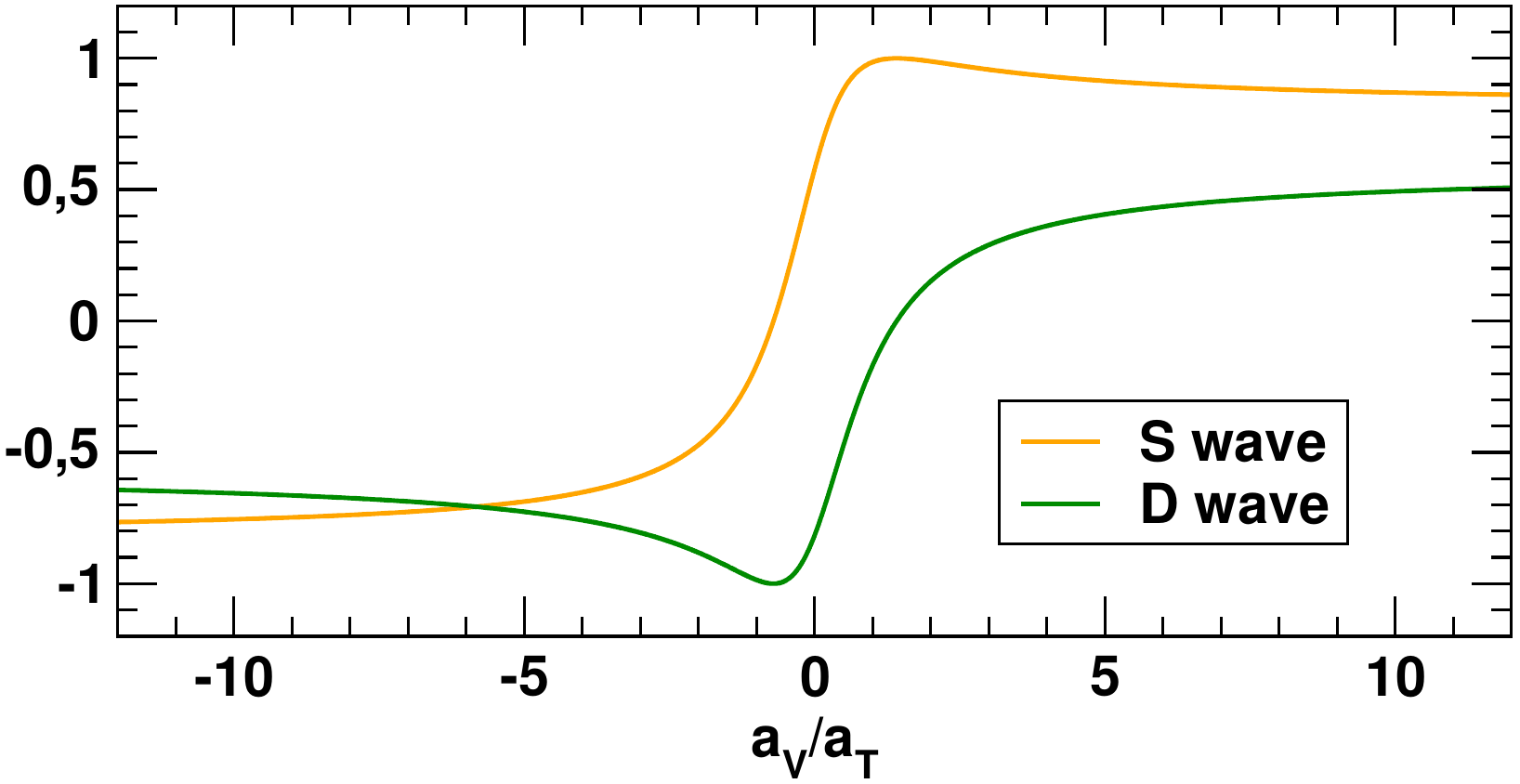}
\caption{Unitary transformation connecting chiral and angular momentum basis.
The expression
$a_V/a_T = \braket{0|J^V_\rho|\rho} / \braket{0|J^T_\rho|\rho}$ denotes
the ratio of interpolator overlap factors with the physical $\rho$ states.}
\label{fig:trafo}
\end{figure}

 In Fig.~\ref{fig:trafo} we
show the dependence of the partial wave content of a state
on the ratio (\ref{ratio}). For example, if the $\rho$ state is a pure $1\;^3S_1$ state,
then two different chiral components 
have to mix in the following way:
\begin{align}
\label{lsj_ground}
\sqrt{2}\ket{\rho_{(0,1)\oplus(1,0)}} + \ket{\rho_{(1/2,1/2)_b}} =
\ket{1;^3S_1}.
\end{align}
In our lattice evaluation for the ground state $\rho$ we will not find the value $\sqrt{2}$ but a value
which is close to it.
%
%
%

\section{Lattice technicalities}
\label{cResolution-scale}

\subsection{Simulation parameters}
We use gauge configurations generously provided by the JLQCD collaboration, see
Ref.~\cite{Aoki:2008tq}. The ensemble consists of 100 configurations of
two-flavor dynamical Overlap fermions. The topological sector is fixed to
$Q_{top}=0$. Lattice size and spacing are $16^3\times 32$ at $a \sim 0.12$ fm.
The pion mass is at $m_\pi = 289(2)$ MeV \cite{Aoki:2008tq}.

We calculate the isovector correlators with extended sources with
different smearing widths of Gaussian type, described below.

\subsection{Resolution scale via smeared sources}
The \textit{vector} current $J_\rho^V$ is conserved, i.e. its coupling $a_V$ to
the physical state should be independent of the scale. The
\textit{pseudotensor} current $J_\rho^T$ is not conserved. Hence, $a_T$ should
depend on the scale where it is measured. Consequently the ratio $a_V/a_T$
should also depend on the scale.

 An
intrinsic resolution scale is set by the lattice spacing $a$. 
If we probe the hadron structure with  the point-like source then the result
should display a structure of a hadron that is obtained at the scale fixed
by the ultraviolet regularization $a$. In principle we could study the $a$-dependence by means of different lattices with different $a$. However, such
a procedure does not allow to measure the structure close to the infrared
region, i.e. at large $a$.

Instead of variing $a$, we can smear
the sources of the quark propagators using different widths $\sigma$. 
Clearly, the smeared source  cannot supply us with the
information about the hadron structure that is sensitive to distances that are smaller than
the smearing width $\sigma$. Consequently the smearing width $\sigma$
defines a scale at which we probe the structure of our hadron.

\begin{figure}
\includegraphics[scale=0.183]{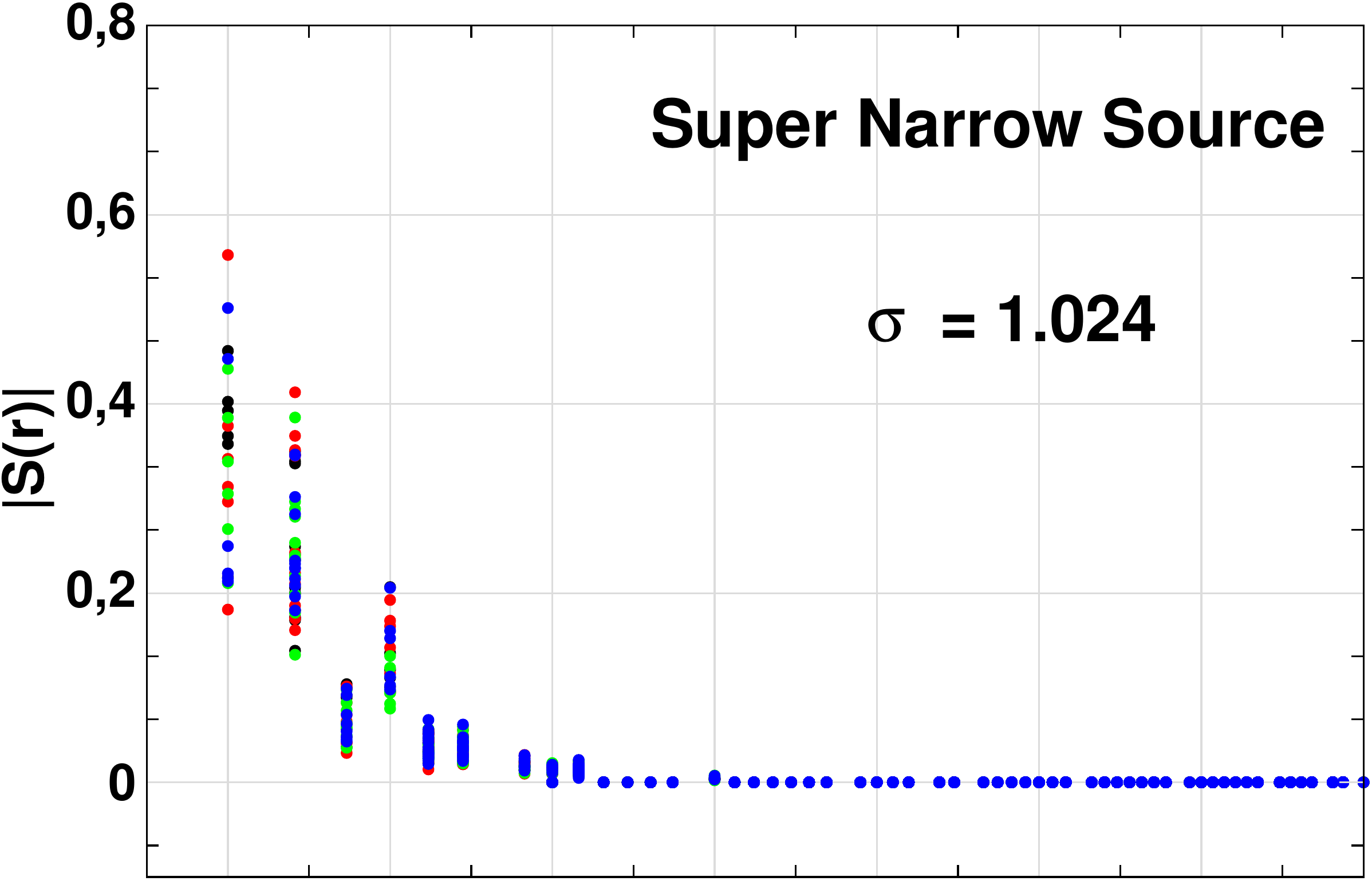}
\includegraphics[scale=0.183]{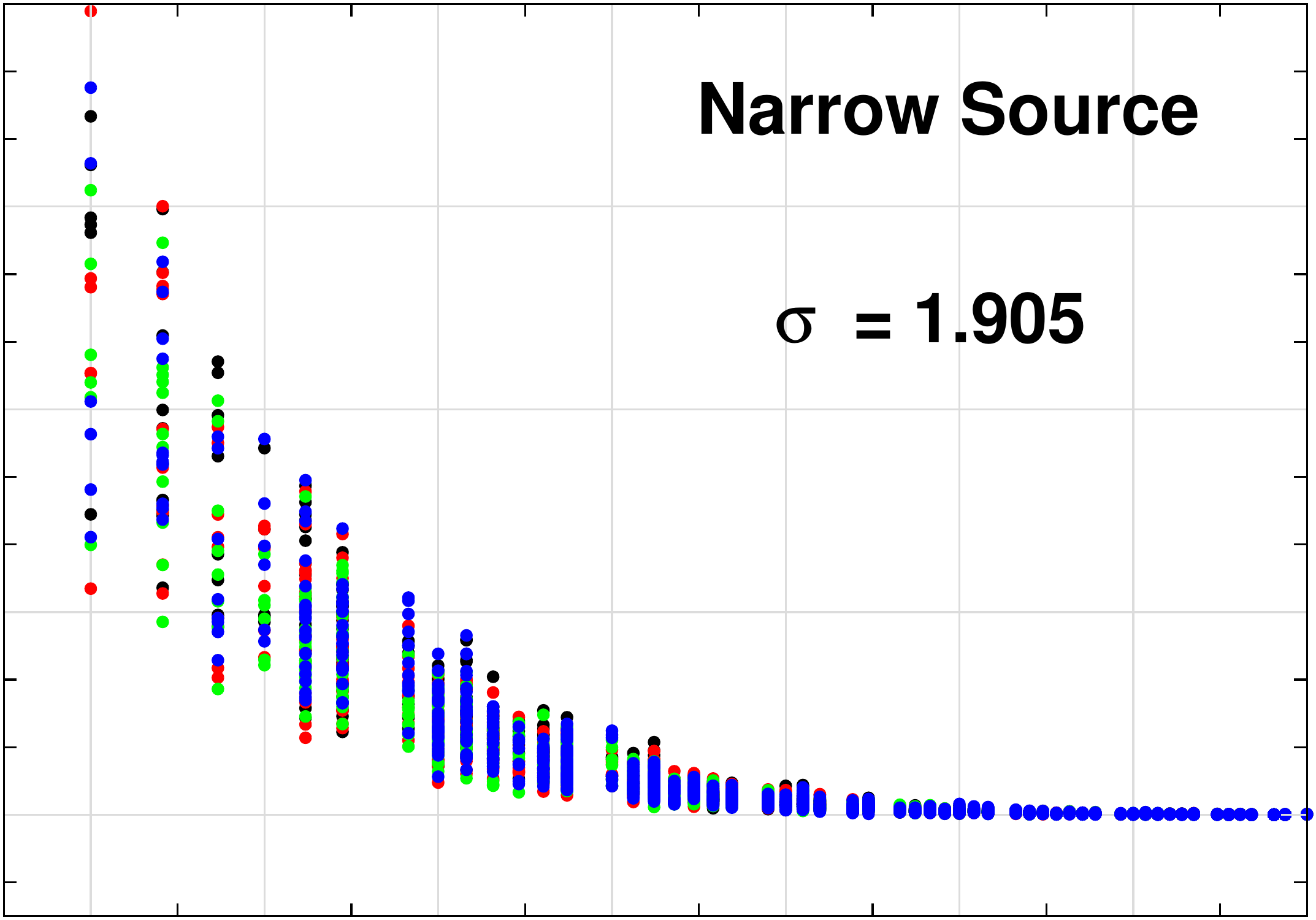}
\includegraphics[scale=0.183]{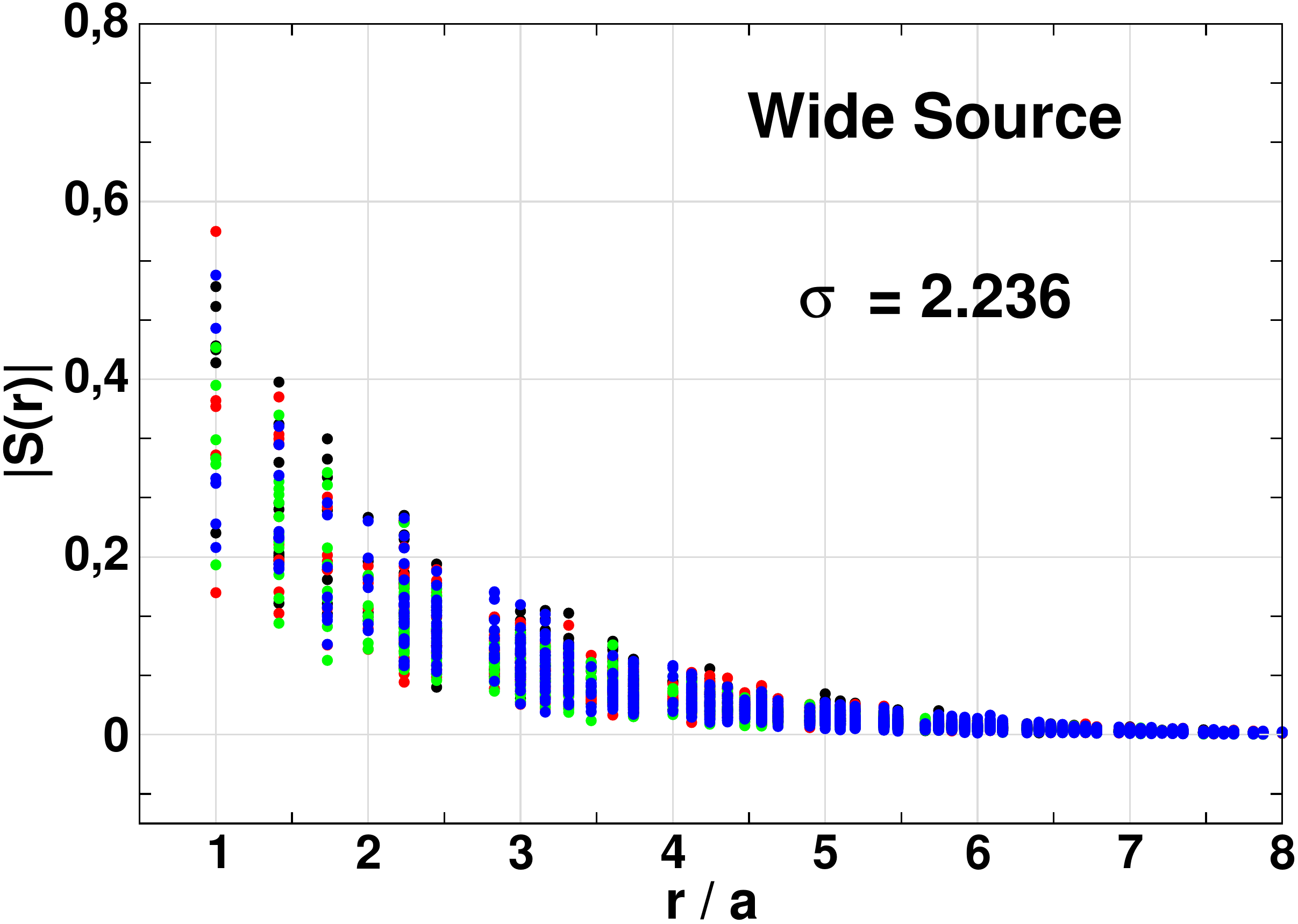}
\includegraphics[scale=0.183]{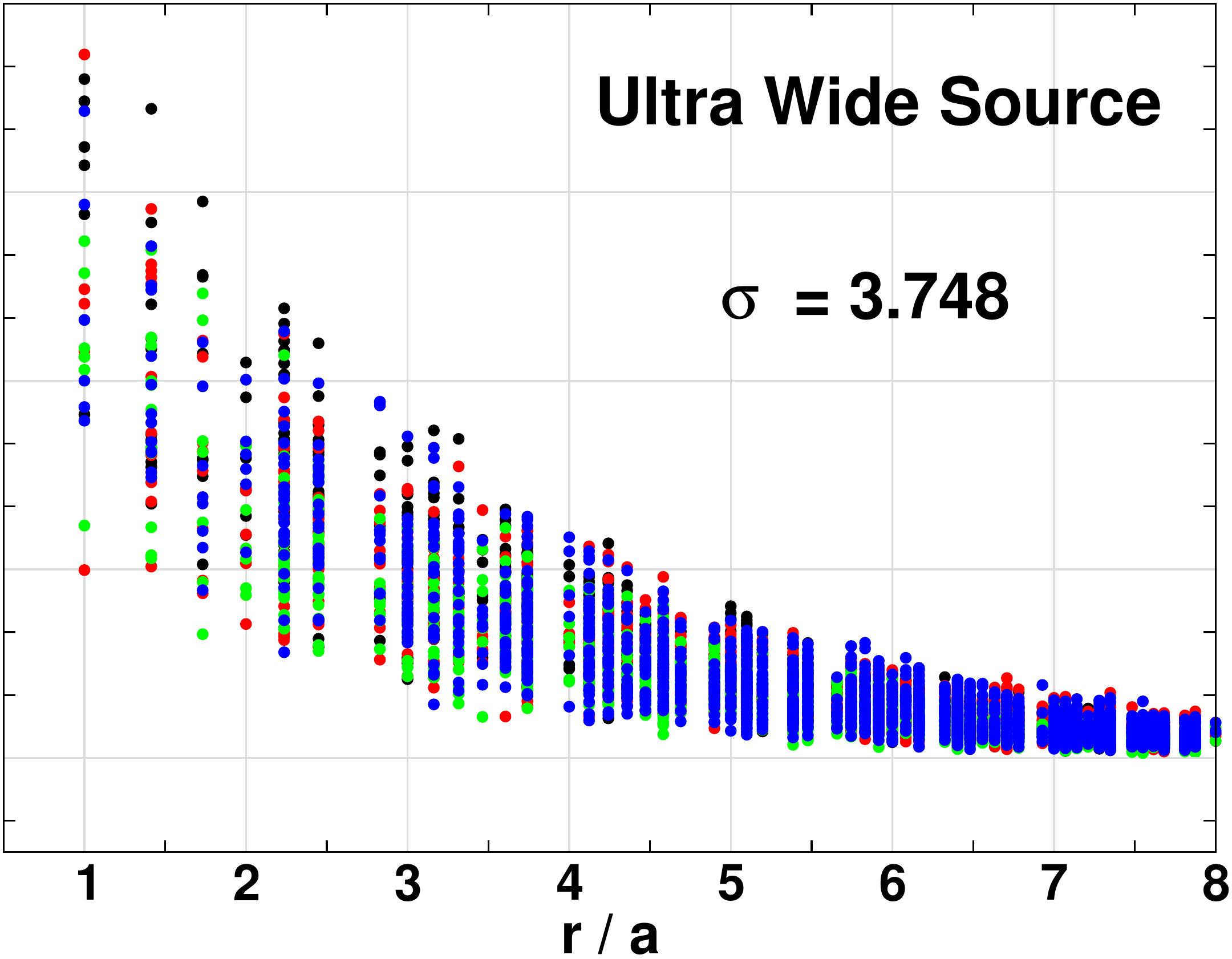}
\caption{Four different source profiles. Different colors correspond to
different gauge configurations.}
\label{fig:sources}
\end{figure}

This is done using
the Gaussian gauge invariant smearing of the source and sink operators. 
For a definition of the Gaussian (Jacobi) smearing we refer to Ref. \cite{Glozman:2009cp} and references therein.
We use four
different smearing widths in this study. The corresponding profiles are given
in Table \ref{tab:sources}. The radius $\sigma$ of a given source $S(x;x_0)$ 
located at $x_0$ is calculated by
\begin{align}
\sigma^2 = \frac{\sum_{\vec{x}}(\vec{x}-\vec{x_0})^2|S(x;x_0)|^2}
                {\sum_{\vec{x}}|S(x;x_0)|^2} \;.
\end{align}
We define the resolution scale as $R=2\sigma a$.
%
%
%
\begin{table}
\center
\caption{
Different Jacobi-smeared sources, generating parameters $\kappa$ and
$N$ \cite{Glozman:2009cp}, widths $\sigma$
and resolution scales $R$.}
\begin{tabular}{ccccccc}
\hline \hline
          & Super-Narrow$\;$& $\;$Narrow $\;$& $\;$Wide   $\;$&$\;$ Ultra-Wide
          															  \\ \hline
$\kappa$  & 0.3             & 0.21           & 0.191          & 0.19  \\
$N$       & 4               & 18             & 41             & 100   \\ \hline
$\sigma$/a& 1.024           & 1.905          & 2.236  		  & 3.748 \\
$\pm$/a   & 0.009           & 0.023  	  	 & 0.068  		  & 0.195 \\ \hline
R/fm      & 0.245           & 0.455  		 & 0.530  		  & 0.890 \\
$\pm$/fm  & 0.003           & 0.005  		 & 0.017  		  & 0.050 \\
\hline \hline
\end{tabular}
\label{tab:sources}
\end{table}
The profiles are pictured in Figure \ref{fig:sources}. The
\textit{Super Narrow} source probes the hadron wave function at 
the resolution $\sim 0.25$ fm and marks the
ultraviolet end of our parameter space. \textit{Narrow} and \textit{Wide} probe
the hadron 
in the mid-momentum region. The \textit{Ultra Wide} source does not resolve details smaller than
$\sim 0.9$ fm and marks our infrared end. In this study it is not reasonable to go
any further in the infrared due to the box size of $\sim 2$ fm.

Note that the gauge configurations remain
untouched throughout the whole process.
%
%
%
%
%
%

\section{Results}
\label{cResults}
As a consistency check we first extract the masses of the $\rho$ states with
our four source profiles and end up with the same results as already found in
Ref.~\cite{Denissenya:2015mqa}.
Masses of mesons, as expected, do not
depend on the resolution scale $R$ and on a choice of a number of smearings  used in
the eigenvalue problem (5).

%
%
%
\begin{figure}[t]
\includegraphics[scale=0.5]{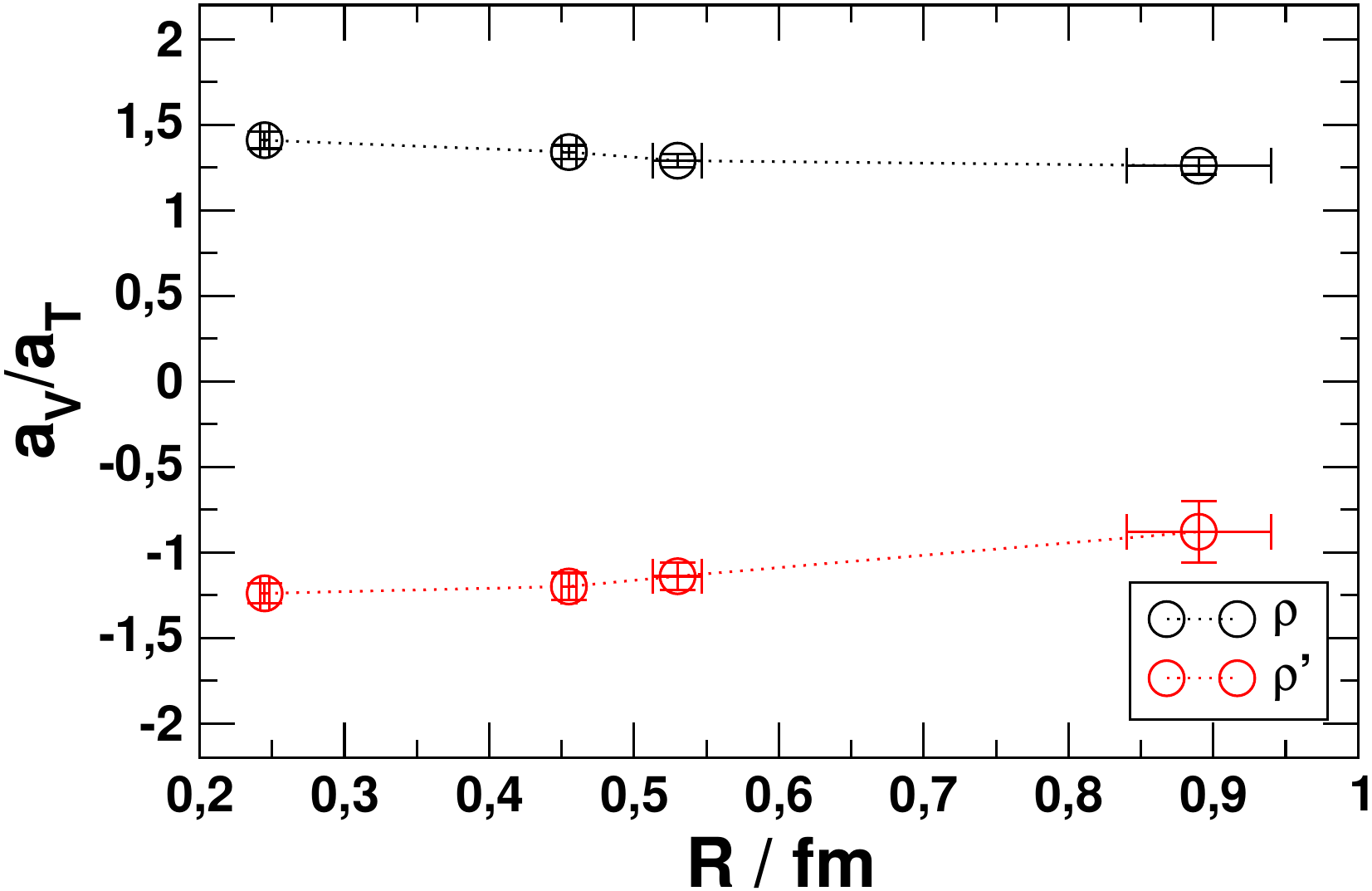}
\caption{$a_V/a_T$ ratio for different resolutions.}
\label{fig:finalplot}
\end{figure}

To study the ratio $a_V/a_T$ at different
resolution scales $R$ we solve the $8 \times 8$ eigenvalue problem (5)-(7)
with operators (1) and (2) and four different smearings. Then using (8)
we exctract the ratio $a_V/a_T$ as a function of $R$.
In  Fig.~\ref{fig:finalplot} we show the ratio $a_V/a_T$ at different
resolution scales $R$.    We find a clear
 $R$-effect on the ratio $a_V/a_T$. For both $\rho$ and $\rho'$ states we see a linear
dependence on the resolution scale between $0.2$ fm and $0.9$ fm.

In the infrared, at the resolution scale
$0.9$ fm, the ratios are given by $a_V/a_T=(1.26\pm 0.05)$ for the ground
state $\rho$ meson and $a_V/a_T = -(0.88\pm 0.18)$ for the first excited
state. Here it is important to note that the ratio for the first excited
state is negative\footnote{We found an error in the data which led to the
conclusions in Refs.~\cite{Glozman:2011gf}, \cite{Glozman:2011nk}. Correcting
this error gives the same result as presented in Fig.~\ref{fig:finalplot}.}.
The ratio $a_V/a_T = 1.26$ for the ground state $\rho$ meson then means that
the chiral representation $(0,1)\oplus(1,0)$ is slightly favoured.
For the first excited state with $a_V/a_T = 0.88$ the $(1/2,1/2)_b$
representation is slightly favoured.  

Using now transformations (\ref{rhovec}),(\ref{rhotensor}) we find:
\begin{align}
\label{equ:rho_wave}
\ket{\rho(770)} =& + \left(0.998 \pm 0.002 \right)   \ket{^3S_1} \\
&- \left(0.05 \;\; \pm 0.025 \right) \ket{^3D_1}\;,\nonumber \\[15pt]
\ket{\rho(1450)} =& - \left(0.106 \pm 0.09\;\; \right) \ket{^3S_1} \\&-
\left(0.994 \pm 0.005 \right) \ket{^3D_1}\;. \nonumber
\end{align}
The ground state $\rho$ is therefore practically a pure $^3S_1$ state, in agreement with the
potential quark model assumption.

The first excited
$\rho$ is, however, a $^3D_1$ state with a very small admixture of a $^3S_1$ wave. 
The latter result is in clear contradiction with the potential constituent quark
model that attributes the first excited state of the $\rho$-meson as a radially
excited $^3S_1$ state.

\section{Effect of low-mode truncation \\on the overlap factors}
\label{cUnbreaking}
We now study the effect of removing the low-lying modes of the Dirac operator
on the ratio $a_V/a_T$. Its effect on the hadron spectrum has been studied
extensively in Refs.~\cite{Denissenya:2014ywa}-\cite{Denissenya:2015woa}. The
Banks-Casher relation connects the low-lying modes of the Dirac operator to the
quark condensate. Hence by removing the lowest eigenmodes we decouple our $\rho$ states from the chiral symmetry breaking dynamics. The procedure of removing the low modes from
the quark propagator ($D^{-1}$ denotes the quark propagator) is given as:
\begin{align}
D^{-1}_k(x,y)=D^{-1}_{FULL}(x,y)-\sum_{i=1}^k \frac{1}{\lambda_i}v_i(x)v_i^\dagger(y).
\end{align}
%
%
%
\begin{figure}[t]
\includegraphics[scale=0.38]{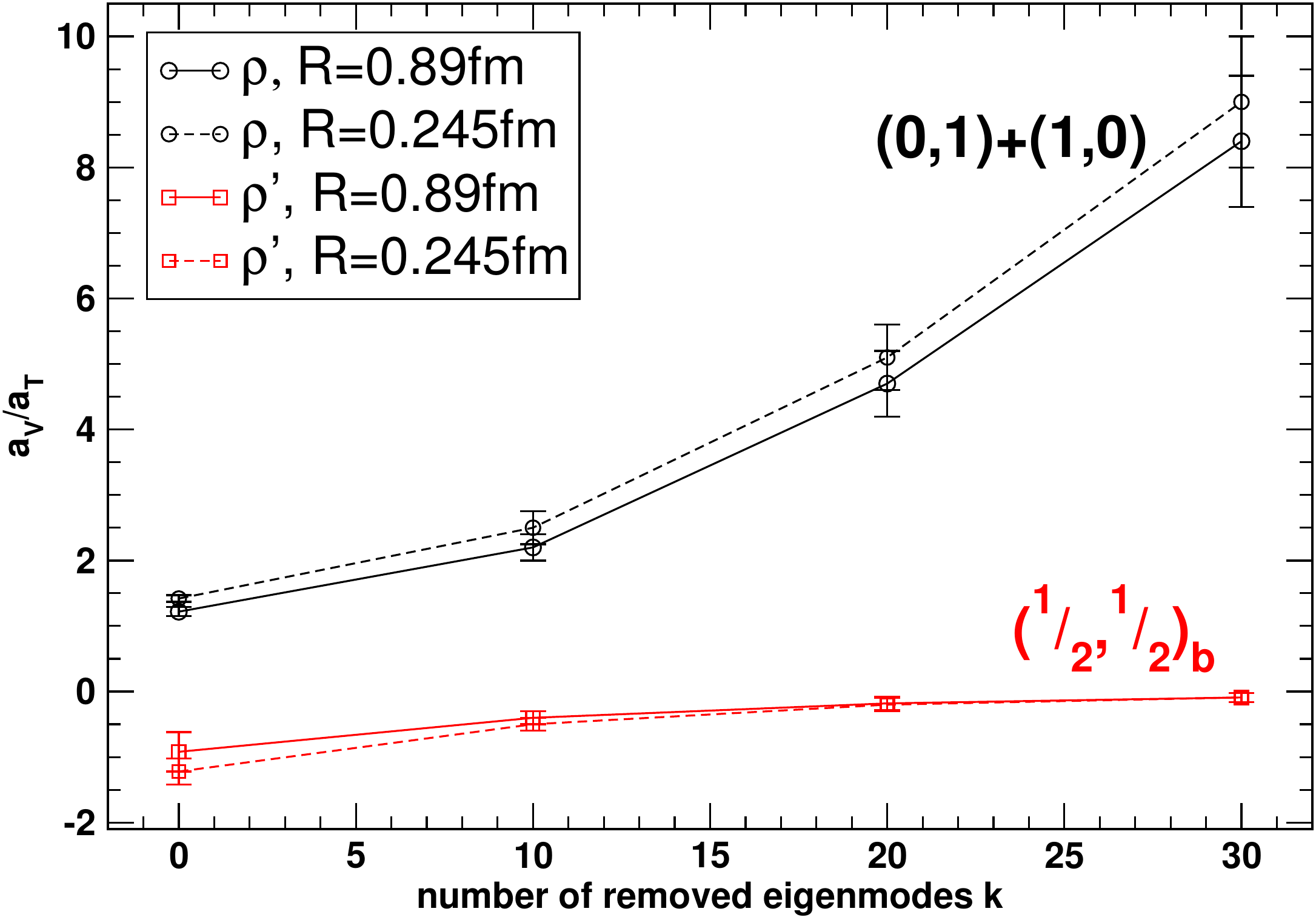}
\caption{Chiral contributions for a low mode truncated system at $R=0.89$ fm
and $R=0.245$ fm.
Here $k=0$ corresponds to the full theory.}
\label{fig:restoration}
\end{figure} 
In Fig.~\ref{fig:restoration} we show the value $a_V/a_T$ for the
ground state $\rho$ and excited $\rho^\prime$ at different resolutions $R$ for
an increasing number of
removed modes. For $k=0$, i.e. the full theory, the mesons are a strong mixture
of both chiral representations. With an increasing number of modes removed the
ground state $\rho$ meson approaches a pure $(0,1)\oplus(1,0)$ state, whereas the
first excited state $\rho^\prime$ becomes a pure $(1/2,1/2)_b$ state. Already
at $k=10$ the states are strongly dominated by one chiral representation: the chiral representations, which are slightly favored for
$k=0$, become dominant for $k\neq 0$.

After removal of $\sim 10-20$ lowest modes both $\rho$ and $\rho'$ get
degenerate, which  reflects a $SU(4)$ symmetry 
\cite{Glozman:2014mka,Glozman:2015qva} of QCD in Euclidean space-time
\cite{Glozman:2015qzf}.
For other recent
studies of this issue see Refs.~\cite{Cohen:2015ekf,Shifman:2016efc}.

\section{Summary and Conclusions}
\label{cConclusions}
In this paper we addressed the issue of the angular momentum
content of the leading quark-antiquark component
of $\rho$ and $\rho(1450)$. In the potential constituent quark model both
states are assumed to be 
$^3S_1$ state. 

We investigated this issue via a lattice simulation with dynamical Overlap
fermions. We studied the ratio of overlap factors of \textit{vector} and
\textit{pseudotensor} interpolators that belong to different chiral representations. We observed this ratio of overlap factors
to be negative for the $\rho(1450)$, which implies that this state is 
 a $^3D_1$ state with only a tiny $^3S_1$ component. The $\rho(770)$ is,
in agreement with the quark model, a $^3S_1$ state. 

Then we studied the effect of removing the low-lying
Dirac eigenmodes on the ratio of overlap factors. The state, which was
identified as the ground state $\rho$ at truncation zero, becomes   the $(1,0)\oplus(0,1)$ state, while
the  $\rho'$ meson becomes a pure $(1/2,1/2)_b$ state. They are both degenerate,
which is a manifestation of the previously found $SU(4)$ symmetry.

\begin{acknowledgments}
We thank the JLQCD collaboration for supplying us with the Overlap gauge
configurations. We also thank M. Denissenya and C.B. Lang for discussions and
help. The calculations have been performed on local clusters at ZID at the
University of Graz and the Graz University of Technology.
Support from the Austrian Science Fund (FWF) through the
grants DK W1203-N16 and P26627-N27 is acknowledged.
\end{acknowledgments}


\end{document}